\documentclass[11pt,preprint]{aastex}

\usepackage{natbib, graphicx, subfigure}
\begin{document}

\title{Keck II Observations of Hemispherical Differences in H$_2$O$_2$ on Europa}

\author{K.P. Hand}
\affil{Jet Propulsion Laboratory, California Institute of Technology, Pasadena, CA 91109}
\email{khand@jpl.nasa.gov}

\author{M.E. Brown}
\affil{Division of Geological and Planetary Sciences, California Institute
of Technology, Pasadena, CA 91125}

\begin{abstract}
We present results from Keck II observations of Europa over four consecutive nights using the near-infrared spectrograph (NIRSPEC).  Spectra were collected in the 3.14--4.0 $\mu$m range, allowing detection and monitoring of the 3.5 $\mu$m feature due to hydrogen peroxide. {\it Galileo} Near-Infrared Spectrometer (NIMS) results first revealed hydrogen peroxide on Europa in the anti-jovian region of the leading hemisphere at an abundance of 0.13$\pm$0.07\% by number relative to water. We find comparable results for the two nights over which we observed the leading hemisphere. Significantly, we observed a small amount of hydrogen peroxide ($\sim$0.04\%) during observations of Europa's anti- and sub-Jovian hemispheres. Almost no hydrogen peroxide was detected during observations of just the trailing hemisphere. We conclude that the {\it Galileo} observations likely represent the maximum hydrogen peroxide concentration, the exception potentially being the cold water ice regions of the poles, which are not readily observable from the ground. Our mapping of the peroxide abundance across Europa requires revisions to previous estimates for Europa's global surface abundance of oxidants and leads to a reduction in the total oxidant delivery expected for the sub-surface ocean, if exchange of surface material with the ocean occurs.
\end{abstract}

 
\section{Introduction}
The interaction of the Jovian magnetosphere with Europa results in radiolytic processing of the icy surface \citep{johnson1990, johnson1997photolysis}. Energetic electrons, ions, and protons dissociate water, yielding OH radicals, which then can recombine forming hydrogen peroxide and other oxidants \citep{cooper2003, johnson2003}.

Importantly, Europa is synchronously locked to Jupiter (Europa rotates on its axis once per 86 hour orbit around Jupiter), and thus Europa's hemispheres maintain the same orientation with respect to Jupiter and the radiation environment of the magnetosphere. Over 75\% of the incident radiation is from electrons \citep{cooper2001} and, due to their small gyroradii and the fast rotation of Jupiter (10 hours) relative to Europa's orbital period, the trailing hemisphere of Europa receives the majority of the radiation. Furthermore, low energy sulfur ions in the plasma torus are preferentially deposited on the trailing hemisphere. This relationship is made evident by the trailing hemisphere `bulls-eye' feature of dark material \citep{paranicas2001}, most likely a radiolytically processed sulfur compound such as hydrated sulfuric acid \citep{carlson1999h2so4} and endogenous magnesium ions that have been radiolytically processed to magnesium sulfate (Brown and Hand, submitted).

 {\it Galileo} Near Infrared Mapping Spectrometer (NIMS) observations of Europa first detected hydrogen peroxide in the anti-Jovian region of the leading hemisphere at an abundance of 0.13+/-0.07\% by number relative to water \citep{carlson1999h2o2}. As shown by \cite{loeffler2006icar}, the equilibrium peroxide concentration for ion-irradiated 80 K and 120 K ice is 0.14\% and 0.1\% by number respectively. Though peroxide has been reported for other regions of Europa, those observations were noisy and poorly quantified  \citep{hansen2008}.
 
 Here we present new results for measurement of the peroxide abundance across all four major hemispheres of Europa (leading, trailing, sub-Jovian, and anti-Jovian) using ground based spectroscopy.

\section{Observations}

Observations of Europa were obtained over four consecutive nights from 17 until 20 September 2011 using NIRCSPEC, the facility near infrared
spectrograph \citep{mclean1998}, on the Keck telescope. We covered a wavelength range from 3.14 - 4.00 $\mu$m in the low resolution mode with R $\sim$ 2000, which is still a factor of $\sim$20 better than the NIMS instrument. We used the 0.76" wide slit, such that for a given observation most of the observable hemisphere of Europa was within the slit. At the time of our observations the 3120 km diameter of Europa subtended 1.028", or approximately 2300 km. While observers of the Galilean satellites often use other Galilean satellites for telluric and solar flux correction, as they are bright and adjacent, we chose, instead, to avoid the possibility of spectral contamination and used HD 9986, a V=6.67 G5V star that was 13 degrees away from Europa at the time of observations. Observations on the target and calibrators were obtained in an ABBA pattern, with a single pointing for Europa consisting of 60 2-second coadds and for the calibrator star 10 1-second coadds. A journal of the observations is shown in Table 1.

The initial data reductions were performed using standard routines that subtract adjacent pairs of images, correct for the curvature in the spatial and spectral dimensions, fit and subtract residual line emission, and optimally extract the remaining spectrum. From the initial extracted spectra, it was clear that the data were affected by variable water vapor absorption and by a line-shape difference between the resolved target and the point source stellar calibrator. To correct for these effects, we first created a grid of model telluric spectra using the online ATRAN \citep{lord1992} tool for SOFIA\footnote{http://atran.sofia.usra.edu/cgi-bin/atran/atran.cgi} spanning a range of air masses and water vapor overburdens, convolved to the resolution of the NIRSPEC data. We divided each stellar spectrum by the model spectra, finding the parameters which minimized the residuals and determined the effects of slight changes in the water vapor burden and airmass for the stellar spectra. We then performed a $\chi^2$ minimization on the calibrated target spectra where we varied the spectral broadening of the target, the precise wavelength offset between the calibrator and target, and the magnitude of small water vapor and airmass corrections. For these bright sources, few atmospheric emission lines are visible in the raw data, so wavelength calibration is performed by matching the model transmission spectra to the uncorrected stellar spectra. The resulting spectra are generally excellent, but with clear uncorrected water vapor residuals shortward of 3.3 $\mu$m.

\section{Results}

The reflectance data for each night was normalized to the maximum reflectance value for the bright, leading hemisphere on the night of the 18th, which showed good agreement with previous albedo measurements for Europa's surface \citep{calvin1995, grundy2007, carlson2009}. 

The surface abundance of hydrogen peroxide, expressed as its percent by number abundance relative to water, was determined by converting reflectance values, $R$, to absorbance, $A = -\ln \left( R / R_{Baseline}  \right)$, where the baseline reflectance used was the maximum reflectance value for the night of the 18th, which was the spectrum representing the purest ice. We then used a second degree least squares polynomial fit to the broad 3 $\mu$m water feature so as to enable removal of the continuum. The peroxide band was integrated to get the total band area from 3.4 to 3.6 $\mu$m (2760 cm$^{-1}$ to 2950 cm$^{-1}$). As part of the polynomial fit we calculated the one standard deviation values above and below the fit and then integrated the corresponding band areas to get 1 $\sigma$ values for the total absorbance. 

To generate peroxide concentrations for each night of observations we used the temperature dependent band strength values from \cite{loeffler2006icar}, and selected A$_{100K} = (5.0\pm0.5)\times10^{-17}$ cm molec$^{-1}$ as a reasonable value for the average hemisphere temperature of 100 K \citep{spencer1999}. Dividing the integrated band strength by the $A$-value yields the total number of H$_2$O$_2$ molecules, which when divided by the number of water molecules for the optical depth yields the concentration relative to water. Following \cite{hudgins1993} and \cite{carlson1999h2o2} we use a 50 $\mu$m grain size and optical depth, and an ice density of $3\times10^{22}$ molecules per square centimeter. Figure 1 shows the averaged spectrum for each night, and shows the polynomial fit and 1 $\sigma$ bounds. The band with the continuum subtracted is shown with the {\it Galileo} NIMS data overlain (x marks).

The leading hemisphere of Europa, as determined by observations with a central Europa longitude ranging from 52$^\circ$ to 62$^\circ$, has a $0.113\pm0.032$ (0.074-0.161) percent by number abundance relative to water, where the range in parenthesis includes the uncertainty in the $A$-value. This range is in close agreement with the NIMS observations, which had an uncertainty of $\pm$0.07. For added comparison we analyzed the {\it Galileo} NIMS data using the more recent band strengths \citep{loeffler2006icar} and found that the NIMS data lead to a percent by number abundance of 0.118\% peroxide relative to water. This value is slightly less than the 0.13\% value reported by \cite{carlson1999h2o2}, but well within the published error bars for that observation, and is in very close agreement with our observations.

The trailing hemisphere of Europa, which is dominated by a darker non-ice material, was found to have a peroxide abundance close to or equal to zero. These observations had a central Europa longitude ranging from 256$^\circ$ to 264$^\circ$ and thus were centered on the apex of the trailing hemisphere. As the spectrum shows, there is only a slight rise above the continuum at 3.5 $\mu$m. The measured concentration range is 0.0 to a maximum of 0.038 (or 0.043 for $A_{100K}=4.5\times10^{-17}$ cm molec$^{-1}$, using the low-end value for the error limits on $A_{100K}$) percent by number abundance, with our best fit value being 0.018 for a 100 K surface.

Observations that included the sub-Jovian (Jupiter facing) and anti-Jovian hemispheres of Europa show peroxide concentrations of approximately 1/3 that of the leading hemisphere, though as shown in Table 1 the longitude ranges for these observations do partially overlap with the leading and trailing hemispheres. The September 19th observations centered on a longitude of 158$^\circ$ to 161$^\circ$, covering the leading to anti-Jovian hemispheres, and indicate a concentration of $0.045\pm0.019$ (0.023-0.072) percent by number abundance. The September 17th observations that centered on a longitude ranging from 315$^\circ$ to 316$^\circ$, and thus covered the trailing to sub-Jovian hemispheres, yield a concentration of $0.042\pm0.041$ (0.0-0.092) at 100 K. 

\section{Discussion}

Our results show a non-uniform distribution of hydrogen peroxide across the surface of Europa. Three key factors likely control the production and abundance of peroxide on Europa's surface: 1) the availability of water molecules, 2) radiation (UV, electrons, and ions all yield peroxide \citep{johnson1997photolysis, johnson2003}), and 3) temperature \citep{moorehudson2000, loeffler2006icar, handcarlson2011h2o2}. Though the trailing hemisphere of Europa has a higher radiation flux, it is also dominated by a strong non-ice component that limits the production and steady-state abundance of hydrogen peroxide. Our results, coupled with water ice abundance maps \citep{grundy2007, carlson2009, brownhand2012}, clearly show that peroxide abundance maps most directly to the availability of water, which is the precursor molecule that provides the OH radical upon dissociation by irradiation \citep{cooper2003}. Though water molecules may present as part of the hydrated sulfate that dominates the trailing hemisphere \citep{carlson2009}, peroxide does not appear to be a dominant product of the sulfate radiolysis.

Ice temperature significantly modulates peroxide production and stability, as the trapping of OH and electrons in the ice matrix \citep{quickenden1991} decreases in efficiency over the $\sim$ 70-130 K temperature range of Europa's surface \citep{spencer1999}, leading to reduced concentrations at higher temperature \citep{handcarlson2011h2o2}. However, variation in band absorption from 80 K to 120 K yields a $\sim$10\% change in concentration  \citep{loeffler2006icar} and thus cannot explain the difference in concentration across hemispheres. Interestingly, comparison with \cite{loeffler2005} indicates that the state of peroxide on Europa is that of dispersed H$_2$O$_2\cdot2$H$_2$O trimers. In warmer regions and on e.g. Ganymede and Callisto - where peroxide has not been observed - the expected state of peroxide is that of the precipitated and crystallized dihydrate. Our ground based observations agree with these lab based measurements for the state of peroxide on Europa.

Given that condensed O$_2$ has been reported as a component in the surface ice of Europa without longitudinal variation \citep{spencer2002}, and that H$_2$O$_2$ is a useful precursor molecule in pathways to O$_2$ production \citep{cooper2003}, the differences in peroxide concentrations may provide some bearing on the origin of Europa's oxygen distribution and production pathways. \cite{teolis2005} note that destruction of hydrogen peroxide alone cannot explain the observed oxygen abundance and thus other pathways are important to consider. We suggest that perhaps \cite{spencer2002} did observe longitudinal variations in O$_2$ that could be associated with H$_2$O$_2$ as a precursor, but given the limited measurements and the sensitivity of the 5770 \AA\ band, they did not have sufficient evidence for such a conclusion. Indeed, close inspection of their Figure 2 reveals that the oxygen absorption on the leading hemisphere is deeper and broader than that of the trailing hemisphere, with a band depth difference of $\sim$10-25\%. Our measurements showing variations in H$_2$O$_2$ across Europa's surface provide a compelling rationale for a more detailed investigation of Europa's O$_2$ abundance and distribution.

Nevertheless, the \cite{spencer2002} observations do show significant O$_2$ on the trailing hemisphere. In light of our results, we propose that O$_2$ may be a more stable radiation product (as is clearly the case in Earth's atmosphere) and thus the low levels of peroxide on the trailing hemisphere do not prohibit a measurable steady state abundance of oxygen on the trailing hemisphere. The high abundance of sulfur on the trailing hemisphere may also serve to scavenge OH radicals to form H$_2$SO$_4$, and SO$_2$ may be an alternative precursor to O$_2$.

Additionally, oxygen can easily migrate across the surface of Europa as it transitions from solid to gas phase with Europa's diurnal temperature cycles \citep{hall1995, cassidy2007}. The leading hemisphere may be the dominant source of O$_2$ production in the ice, but warming releases that oxygen to the exosphere, where it then migrates and recondenses depending on the diurnal thermal pulse. This scenario is further aided by the fact that UV photolysis is enhanced during the day, and photolysis of peroxide yields OH and possibly alternative pathways for O$_2$ formation \citep{grunewald1986, johnson2003}.

Finally, we note that the radiolytic production of surface oxidants has long been of interest in the context of Europa's subsurface ocean chemistry \citep{gaidos1999, chyba2000energy, chybahand2001}. If Europa's oxidant laden surface ice mixes with the ocean water then radiolysis could be a key mechanism for maintaining a chemically-rich and potentially habitable ocean \citep{chyba2000energy, hand2009}. Previous estimates all assumed a globally uniform layer of peroxide within the ice layer and calculated delivery rates based the NIMS concentration of 0.13\% relative to water. 

Our new results indicate that only the most ice-rich regions of Europa reach the concentrations measured by NIMS. The trailing hemisphere concentration is nearly an order of magnitude lower than the leading hemisphere and the sub- and anti-Jovian hemispheres are down by a factor of $\sim$3 relative to the leading hemisphere. As a result, the average global surface abundance of peroxide in the surface ice of Europa may be better represented by the sub- and anti-Jovian hemisphere concentrations of $\sim$0.044\%. This reduces the low-end estimate of \cite{chybahand2001} from $\sim10^{9}$ moles per year peroxide delivered to the ocean to $\sim10^{8}$ moles per year delivered. Compared to models for seafloor production of reductants, such as methane and hydrogen sulfide, which yield $\sim3\times10^9$ moles per year delivered to the ocean, it appears that our new results for peroxide on Europa could lead to an ocean limited by oxidant availability. This conclusion also depends strongly on the global geographic distribution of O$_2$, which may have concentrations significantly larger than peroxide \citep{hand2007energy}.

\acknowledgements 
This research has been supported by grant
NNX09AB49G from the NASA Planetary Astronomy program and by the NASA Astrobiology Institute 'Icy Worlds' node at JPL/Caltech. The authors thank Robert W. Carlson for helpful discussions.


\begin{thebibliography}{33}
\expandafter\ifx\csname natexlab\endcsname\relax\def\natexlab#1{#1}\fi

\bibitem[{Bain \& Gigu{\`e}re(1955)}]{bain1955}
Bain, O. \& Gigu{\`e}re, P. 1955, Canadian Journal of Chemistry, 33, 527

\bibitem[{Brown \& Hand(submitted)}]{brownhand2012}
Brown, M. \& Hand, K. submitted, Science

\bibitem[{Calvin {et~al.}(1995)Calvin, Clark, Brown, \& Spencer}]{calvin1995}
Calvin, W., Clark, R., Brown, R., \& Spencer, J. 1995, Journal of Geophysical
  Research, 100, 19041

\bibitem[{Carlson {et~al.}(1999{\natexlab{a}})Carlson, Anderson, Johnson,
  Smythe, Hendrix, Barth, Soderblom, Hansen, McCord, Dalton,
  {et~al.}}]{carlson1999h2o2}
Carlson, R., Anderson, M., Johnson, R., Smythe, W., Hendrix, A., Barth, C.,
  Soderblom, L., Hansen, G., McCord, T., Dalton, J., {et~al.}
  1999{\natexlab{a}}, Science, 283, 2062

\bibitem[{Carlson {et~al.}(2009)Carlson, Calvin, Dalton~III, Hansen, Hudson,
  Johnson, McCord, \& Moore}]{carlson2009}
Carlson, R., Calvin, W., Dalton~III, J., Hansen, G., Hudson, R., Johnson, R.,
  McCord, T., \& Moore, M. Europa, ed. R.~Pappalardo, W.~B. McKinnon, \&
  K.~Khurana (University of Arizona Press), 283--327

\bibitem[{Carlson {et~al.}(1999{\natexlab{b}})Carlson, Johnson, \&
  Anderson}]{carlson1999h2so4}
Carlson, R., Johnson, R., \& Anderson, M. 1999{\natexlab{b}}, Science, 286, 97

\bibitem[{Cassidy {et~al.}(2007)Cassidy, Johnson, McGrath, Wong, \&
  Cooper}]{cassidy2007}
Cassidy, T., Johnson, R., McGrath, M., Wong, M., \& Cooper, J. 2007, Icarus,
  191, 755

\bibitem[{Chyba(2000)}]{chyba2000energy}
Chyba, C. 2000, Nature, 403, 381

\bibitem[{Chyba \& Hand(2001)}]{chybahand2001}
Chyba, C. \& Hand, K. 2001, Science, 292, 2026

\bibitem[{Cooper {et~al.}(2001)Cooper, Johnson, Mauk, Garrett, \&
  Gehrels}]{cooper2001}
Cooper, J., Johnson, R., Mauk, B., Garrett, H., \& Gehrels, N. 2001, Icarus,
  149, 133

\bibitem[{Cooper {et~al.}(2003)Cooper, Johnson, \& Quickenden}]{cooper2003}
Cooper, P., Johnson, R., \& Quickenden, T. 2003, Icarus, 166, 444

\bibitem[{Gaidos {et~al.}(1999)Gaidos, Nealson, \& Kirschvink}]{gaidos1999}
Gaidos, E., Nealson, K., \& Kirschvink, J. 1999, Science, 284, 1631

\bibitem[{Grundy {et~al.}(2007)Grundy, Buratti, Cheng, Emery, Lunsford,
  McKinnon, Moore, Newman, Olkin, Reuter, {et~al.}}]{grundy2007}
Grundy, W., Buratti, B., Cheng, A., Emery, J., Lunsford, A., McKinnon, W.,
  Moore, J., Newman, S., Olkin, C., Reuter, D., {et~al.} 2007, Science, 318,
  234

\bibitem[{Grunewald {et~al.}(1986)Grunewald, Gericke, \& Comes}]{grunewald1986}
Grunewald, A., Gericke, K., \& Comes, F. 1986, Chemical Physics Letters, 132,
  121

\bibitem[{Hall {et~al.}(1995)Hall, Strobel, Feldman, McGrath, \&
  Weaver}]{hall1995}
Hall, D., Strobel, D., Feldman, P., McGrath, M., \& Weaver, H. 1995, Nature,
  373, 677

\bibitem[{Hand \& Carlson(2011)}]{handcarlson2011h2o2}
Hand, K. \& Carlson, R. 2011, Icarus, doi:10.1016/j.icarus.2011.06.031

\bibitem[{Hand {et~al.}(2007)Hand, Carlson, \& Chyba}]{hand2007energy}
Hand, K., Carlson, R., \& Chyba, C. 2007, Astrobiology, 7, 1006

\bibitem[{Hand {et~al.}(2009)Hand, Chyba, Priscu, Carlson, \&
  Nealson}]{hand2009}
Hand, K., Chyba, C., Priscu, J., Carlson, R., \& Nealson, K. Europa, ed.
  R.~Pappalardo, W.~B. McKinnon, \& K.~Khurana (University of Arizona Press),
  589--629

\bibitem[{Hansen \& McCord(2008)}]{hansen2008}
Hansen, G. \& McCord, T. 2008, Geophysical Research Letters, 35, L01202

\bibitem[{Hudgins {et~al.}(1993)Hudgins, Sandford, Allamandola, \&
  Tielens}]{hudgins1993}
Hudgins, D., Sandford, S., Allamandola, L., \& Tielens, A. 1993, The
  Astrophysical Journal Supplement Series, 86, 713

\bibitem[{Johnson(1990)}]{johnson1990}
Johnson, R. 1990, {Energetic Charged-Particle Interactions with Atmospheres and
  Surfaces} (Springer-Verlag, New York), 232

\bibitem[{Johnson \& Quickenden(1997)}]{johnson1997photolysis}
Johnson, R. \& Quickenden, T. 1997, J. Geophys. Res, 102, 10985

\bibitem[{Johnson {et~al.}(2003)Johnson, Quickenden, Cooper, McKinley, \&
  Freeman}]{johnson2003}
Johnson, R., Quickenden, T., Cooper, P., McKinley, A., \& Freeman, C. 2003,
  Astrobiology, 3, 823

\bibitem[{Loeffler \& Baragiola(2005)}]{loeffler2005}
Loeffler, M. \& Baragiola, R. 2005, Geophys. Res. Lett, 32

\bibitem[{Loeffler {et~al.}(2006)Loeffler, Raut, Vidal, Baragiola, \&
  Carlson}]{loeffler2006icar}
Loeffler, M., Raut, U., Vidal, R., Baragiola, R., \& Carlson, R. 2006, Icarus,
  180, 265

\bibitem[{Lord(1992)}]{lord1992}
Lord, S. 1992, Ames Research Center, Moffett Field, CA

\bibitem[{McLean {et~al.}(1998)McLean, Becklin, Bendiksen, Brims, Canfield,
  Figer, Graham, Hare, Lacayanga, Larkin, {et~al.}}]{mclean1998}
McLean, I., Becklin, E., Bendiksen, O., Brims, G., Canfield, J., Figer, D.,
  Graham, J., Hare, J., Lacayanga, F., Larkin, J., {et~al.} 1998, in
  Astronomical Telescopes \& Instrumentation, International Society for Optics
  and Photonics, 566--578

\bibitem[{Moore \& Hudson(2000)}]{moorehudson2000}
Moore, M. \& Hudson, R. 2000, Icarus, 145, 282

\bibitem[{Paranicas {et~al.}(2001)Paranicas, Carlson, \&
  Johnson}]{paranicas2001}
Paranicas, C., Carlson, R., \& Johnson, R. 2001, Geophysical Research Letters,
  28, 673

\bibitem[{Quickenden {et~al.}(1991)Quickenden, Matich, Bakker, Freeman, \&
  Sangster}]{quickenden1991}
Quickenden, T., Matich, A., Bakker, M., Freeman, C., \& Sangster, D. 1991, The
  Journal of Chemical Physics, 95, 8843

\bibitem[{Spencer \& Calvin(2002)}]{spencer2002}
Spencer, J. \& Calvin, W. 2002, Astron. J, 124, 3400

\bibitem[{Spencer {et~al.}(1999)Spencer, Tamppari, Martin, \&
  Travis}]{spencer1999}
Spencer, J., Tamppari, L., Martin, T., \& Travis, L. 1999, Science, 284, 1514

\bibitem[{Teolis {et~al.}(2005)Teolis, Vidal, Shi, \& Baragiola}]{teolis2005}
Teolis, B., Vidal, R., Shi, J., \& Baragiola, R. 2005, Physical Review B, 72,
  245422

\end{thebibliography}

\clearpage

\begin{deluxetable}{llcccc}
\tablecaption{Summary of Europa observations.}
\tablehead{\colhead{date}&\colhead{target}&\colhead{time}&\colhead{airmass} &\colhead{longitude} &\colhead{exp. time} \\
\colhead{(UT)}&\colhead{}&\colhead{start/end}&\colhead{start/end}&\colhead{start/end}&\colhead{(sec)}}
\startdata
2011 Sep 17    &Europa --    &12:45/12:54    & 1.01/1.01    &315/316    & 840 \\
       &\ \ \ \ sub-Jup iter    &    &     &    & \\
       &HD 9866    &13:56    & 1.10    &    & 40 \\
2011 Sep 18    &HD 9866    &11:40    & 1.02    &    & 40 \\
       &Europa --     &11:47/13:58    & 1.06/1.02    & 52/62    & 6240 \\
       &\ \ \ \ leading        &    &     &     & \\
2011 Sep 19    &HD 9866    &11:25    & 1.02    &    & \\
       &Europa --    &12:50/13:20  & 1.01/1.01    &158/161    & 1440 \\
       &\ \ \ \ anti-Jupiter        &    &     &    & \\
2011 Sep 20    &HD 9866    &11:48    & 1.04    &    & 40 \\
       &Europa --     &11:55/13:57  & 1.04/1.04  &256/264      & 5760 \\
       &\ \ \ \ trailing        &    &   &    & \\
\enddata
\end{deluxetable}

\begin{figure}[ht!]
     \begin{center}
        \subfigure[]{
            \label{fig:first}
            \includegraphics[width=0.5\textwidth]{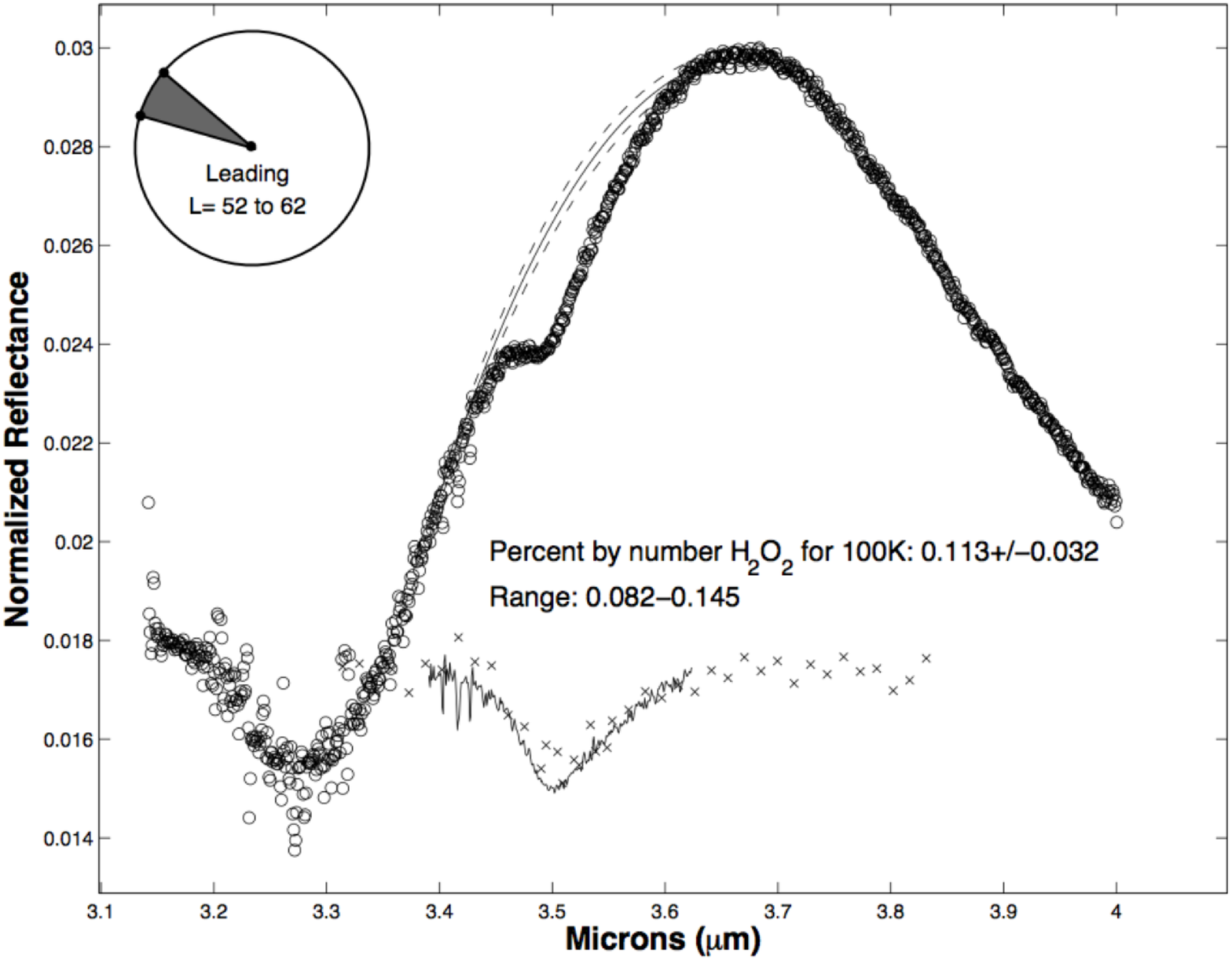}
        }%
        \subfigure[]{
           \label{fig:second}
           \includegraphics[width=0.5\textwidth]{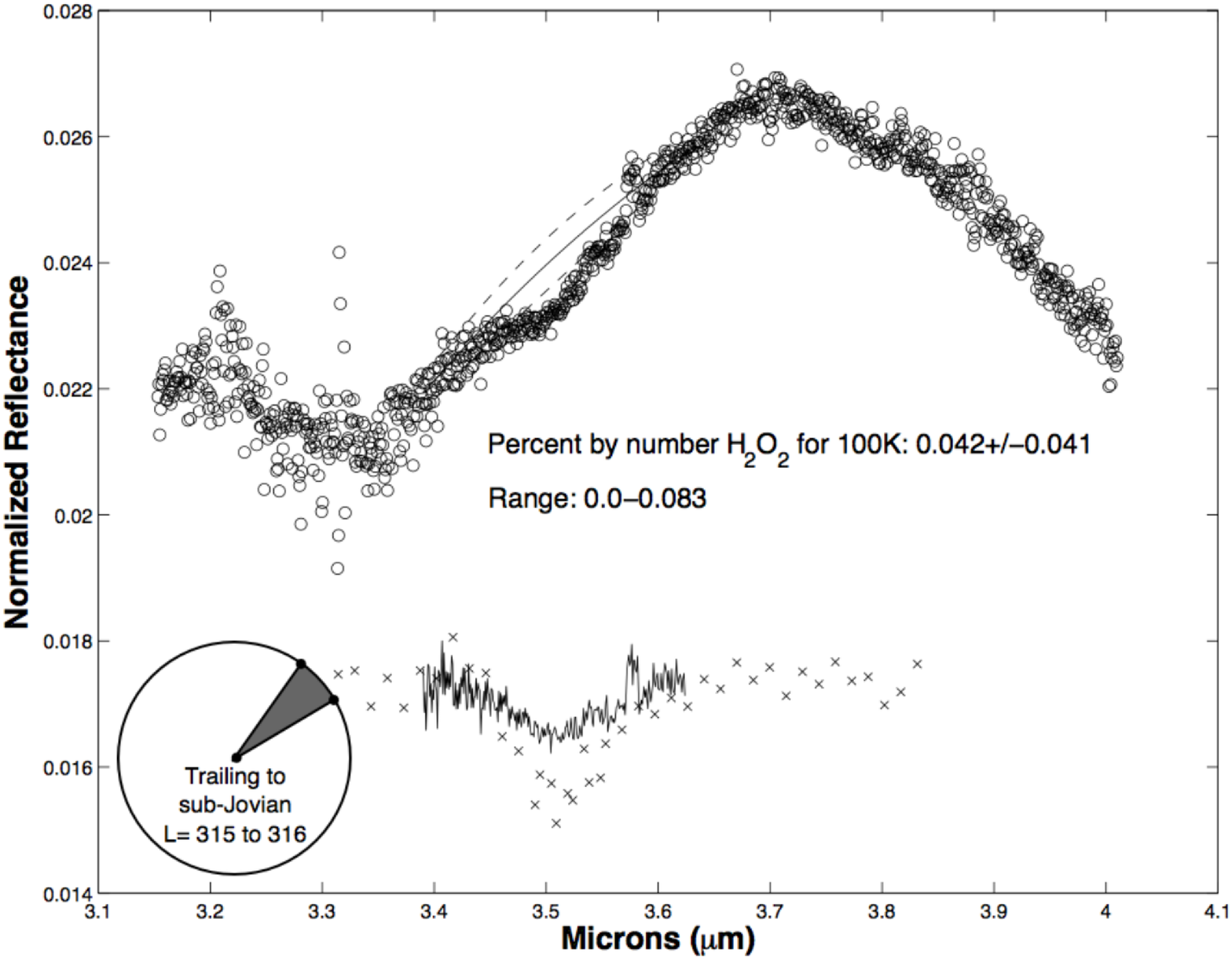}
        }\\ 
        \subfigure[]{
            \label{fig:third}
            \includegraphics[width=0.5\textwidth]{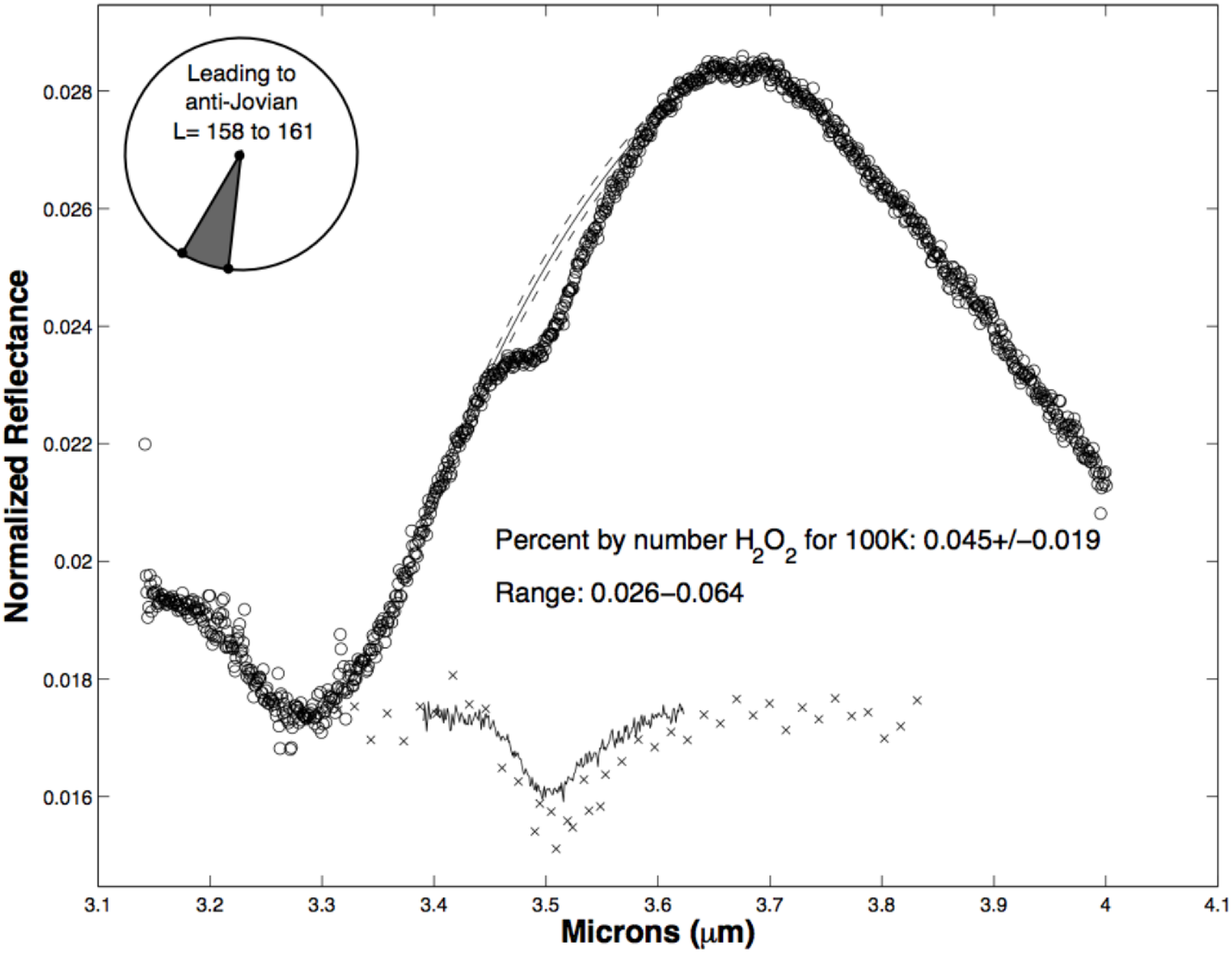}
        }%
        \subfigure[]{
            \label{fig:fourth}
            \includegraphics[width=0.5\textwidth]{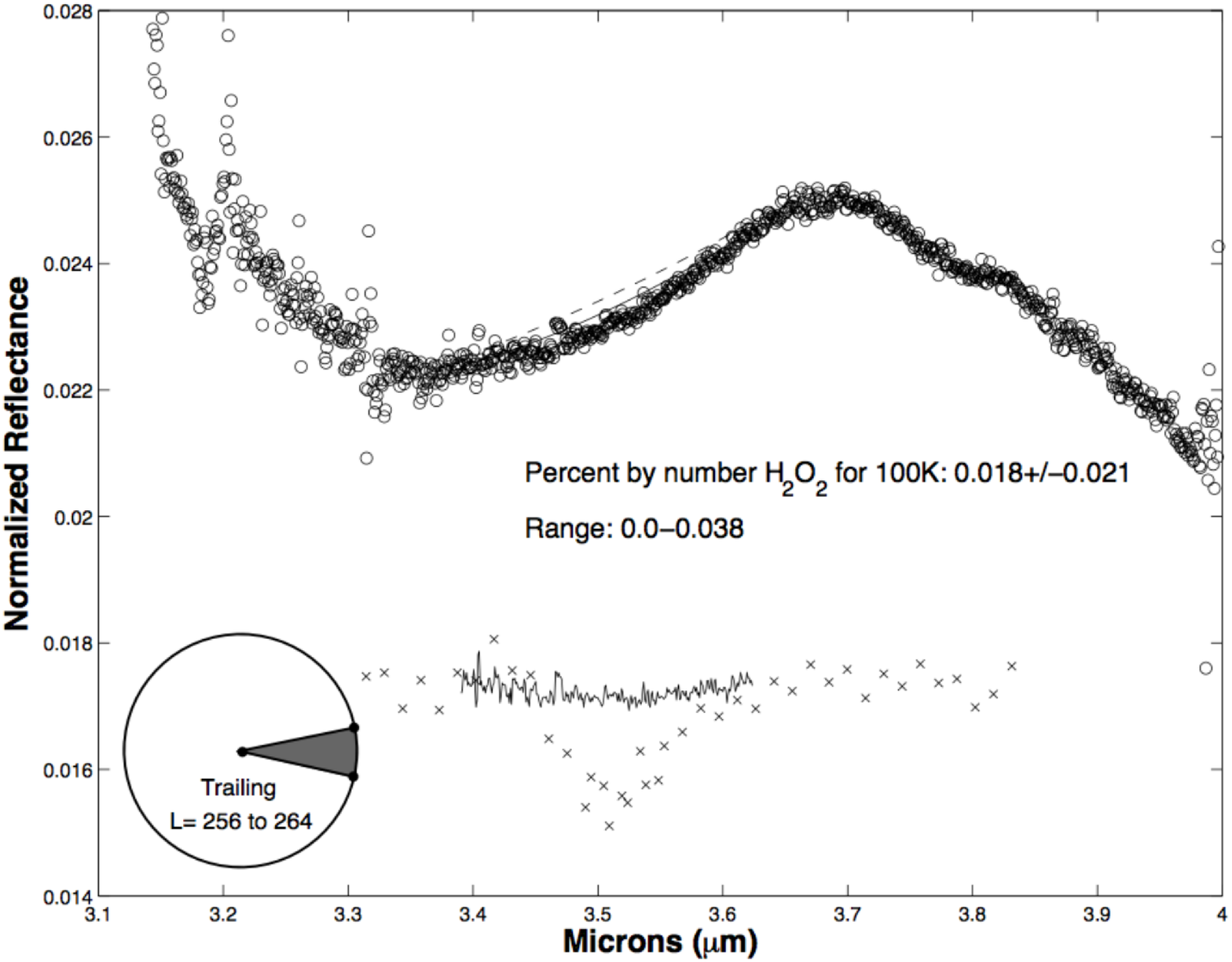}
        }%
    \end{center}
    \caption{%
        Averaged spectrum for each night ($\sim$ 6 hrs) showing the polynomial fit (solid line through the data) and 1 $\sigma$ bounds (dashed lines). Also shown is the hydrogen peroxide band with the water continuum subtracted and the {\it Galileo} NIMS data overlain (x marks). The diagram within each spectrum shows a top-down view of Europa's orbit, with Jupiter at the center and the Earth toward the bottom of the page. The grey wedge shows the $\sim$25$^\circ$ region subtended by Europa as it moves in its orbit over the 6 hour observation period for each night. L is the central Europa longitude of the observations.
     }%
   \label{fig:subfigures}
\end{figure}

\end{document}